# Energetic, structural and electronic features of Sn-, Ga-, O-based defect complexes in cubic $In_2O_3$

*Alexandr I. Cocemasov[1,‡], Vladimir I. Brinzari[1,‡], Denis L. Nika[1,*]*

[1]E. Pokatilov Laboratory of Physics and Engineering of Nanomaterials, Department of Theoretical Physics, Moldova State University, Chisinau, MD-2009, Republic of Moldova

**Abstract**

Defect energy formation, lattice distortions and electronic structure of cubic $In_2O_3$ with Sn, Ga and O impurities were theoretically investigated using density functional theory. Different types of point defects, consisting of 1 to 4 atoms of Sn, Ga and O in both substitutional and interstitial (structural vacancy) positions, were examined. It was demonstrated, that formation of substitutional Ga and Sn defects are spontaneous, while formation of interstitial defects requires an activation energy. The donor-like behavior of interstitial Ga defects with splitting of conduction band into two subbands with light and heavy electrons, respectively, was revealed. Contrarily, interstitial O defects demonstrate acceptor-like behavior with the formation of acceptor levels or subbands inside the band gap. The obtained results are important for an accurate description of transport phenomena in $In_2O_3$ with substitutional and interstitial defects.

## 1. Introduction

Indium oxide is one of the basic transparent semiconducting metal oxides, whose advantages and unique properties have been demonstrated in widespread applications in micro- and optoelectronics, chemical sensors, catalysis, photovoltaics, thermoelectric devices, etc. [1-3]. In spite of rather intensive fundamental studies some aspects of its physico-chemical, electronic and magnetic behaviour remain unclear and their understanding may serve for further developments in aforementioned and possibly new areas of application. The crystallographic complexity of cubic $In_2O_3$ lattice impedes studies and interpretation of its physical properties. Even primitive cell consists of 8 formula units, i.e. 40 atoms.

One of the characteristic features of $In_2O_3$ lattice is an existence of structural vacancies (SVs) which form a kind of voids in the crystal lattice. The proportion of such vacancies in the unit cell is equal to the number of formula units. Additionally, the presence of structural vacancies results in distortions of oxygen in the 1-st shell surrounding indium atoms. These lattice sites differ from ordinary interstitial sites because they can be considered as a continuation of lattice nodes and are placed at positions of so-called "missed" anions. Therefore, it seems that the nature and point defect formation due to the impurity atoms embedded in SVs should be different in comparison with ordinary interstitial sites. Gallium is one of the interesting and prominent candidates for doping of binary In-O, ternary In-X-O and quaternary In-X-Y-O oxide materials where X, Y=Zn or Sn [4, 5]. Experimental studies [6] of $In_2O_3$:Ga thick films with large grains in µm scale have showed a strong increase of electrical conductivity versus Ga additive increase (more than 1 order of magnitude at 8 at. % Ga). Within the concept of isovalent impurity substitution by III group element it is difficult to explain such behavior and donor-like effect of Ga atom. Moreover, a weak monotonic decrease with Ga concentration of fundamental gap was observed.

Thermal conductivity also demonstrates drastic decrease with Ga and Sn doping in the same concentration range [6, 7] and there is a lack of unambiguous explanation of this drop. Strong dependence of thermal conductivity on the defect concentration is crucial for phonon engineering, i.e. tuning of thermal and/or electrical conduction of materials via modification of their phonon properties [8-10].

In this paper we theoretically investigate the electronic band structures, formation energies, partial charges and bond configurations of $In_2O_3$ structure with different point defects conformed by Sn, Ga and O atoms that include both In lattice nodes (substitutional defects) and also structural vacancies (interstitial defects) as a possible sites for an atom's inclusion. Hereafter we will refer to interstitial atoms as those located in structural vacancies. Employing density functional theory (DFT), we determine the geometrical and energetic (formation) parameters of these effects and their impact on the unit cell structure. We also analyze donor or acceptor behavior of these effects as well as defect-induced modification of the electronic band structure, Bader partial charges [11] and electronic density of states (DOS).

The reminder of the paper is arranged as follows. In Section 2 we describe our computational model of $In_2O_3$ with various point defects. Discussions of defect energetics and bonds configuration are given in Section 3.1 and 3.2 respectively. Section 3.3 describes key-features of electronic band structure in $In_2O_3$ with point defects conformed by Sn, Ga and O atoms. Conclusions are given in Section 4.

## 2. Computational Details

All electronic calculations were performed within density functional theory formalism as implemented in the Quantum Espresso first-principles simulation package [12, 13]. The generalized gradient approximation (GGA) for exchange-correlation functional of Perdew, Burke and Ernzerhof (PBEsol) [14] was used. The plane wave cutoff energy of 650 eV and 3x3x3 *k*-point mesh of the Monkhorst-Pack type [15] were found sufficient to converge the total energy and atomic coordinates. For local density of states (LDOS) and band structure calculations a finer 8x8x8 *k*-point grid was employed. Using the supercell approach the defect-free $In_2O_3$ with bixbyite crystal structure (space group Ia-3) and with various point defects containing Sn, Ga and O impurity atoms were modeled. A 40-atoms primitive cell was chosen as a defect-free reference. Following the Wyckoff notation [16] the reference cell contains 4 indium atoms in position *b*, 12 indium atoms in position *d*, 16 oxygen atoms in position *e* and 8 SVs in position *c*. Both In(b) and In(d) atoms are surrounded by six oxygen atoms, while all O atoms are four-fold coordinated with indium. The position of *b*, *d*, *e*, and *c* sites in $In_2O_3$ lattice is shown in Figure 1. It is seen that the *c*-sites provide the natural space to accommodate the interstitial atoms.

**<Figure 1>**

The $In_2O_3$ with various point defects was simulated by adding (removing) a certain number of In, O, Sn or Ga atoms to (from) the defect-free cell. The characteristic sites constituting the defects are marked in Figure 1 as follows: $b_0$ and $d_0$ are *b*- and *d*-sites for substitutional atoms, $c_0$, $c_1$, $c_2$ - *c*-sites for interstitial atoms, $e_0$ – *e*-site for an oxygen vacancy. All cells were structurally relaxed until forces acting on the ions became below 0.01 eV/Å and internal stress decreased below 0.005 GPa. For pure $In_2O_3$ the equilibrium lattice constant was found to be $a$=10.157 Å, only 0.4% larger than the experimental value 10.117 Å [17]. The calculated volumetric density was 7.04 g/cm$^3$ that is about 2% less than experimental value 7.18 g/cm$^3$[18].

## 3. Results and Discussion

### 3.1. Point defect energetics

Following the Refs. [19, 20] the formation energy (FE) of a neutral point defect $d$ can be obtained from DFT total energy calculations as:

$$E^d_{form} = E^d_{tot} - E^0_{tot} - \sum_i n^d_i \mu_i, \quad (1)$$

where $E^d_{tot}$ is the total energy of a supercell calculation containing the defect $d$, $E^0_{tot}$ is the total energy of a defect-free supercell calculation, $n^d_i$ is the number of atoms of type $i$ that have been added to ($n^d_i > 0$) or removed from ($n^d_i < 0$) the supercell to form the defect, $\mu_i$ is the chemical potential of $i$-th atom type. For complex defects containing several impurity atoms we have presented formation energy per one atom.

Within the thermodynamic approach chemical potentials of In, Sn, Ga and O can be expressed through thermochemical heats of phase formations for $In_2O_3$, $SnO_2$, $Ga_2O_3$ and $O_2$ using the following expressions:

$$2\mu_{In} + 3\mu_O = \Delta H_{In_2O_3}, \quad (2)$$

$$\mu_{Sn} + 2\mu_O = \Delta H_{SnO_2}, \quad (3)$$

$$2\mu_{Ga} + 3\mu_O = \Delta H_{Ga_2O_3}, \quad (4)$$

$$2\mu_O = \Delta H_{O_2}, \quad (5)$$

where $\Delta H_{In_2O_3}$, $\Delta H_{SnO_2}$, $\Delta H_{Ga_2O_3}$ and $\Delta H_{O_2}$ are the respective heats of formation. Note, for exothermic reactions these values are negative. Each of Eqs. (2-5) describes only the part of total thermochemical reaction - the formation process of the corresponding compound. This means that the rate of reverse process (decomposition) is too small and can be excluded from consideration. Otherwise, the establishment of equilibrium would lead to changes in the basic thermodynamic parameters of the system, including chemical potentials. Therefore, the sum of chemical potential values (left side of Eqs. (2-5)) determines the transition boundary between the formation/decomposition of secondary phases only. The appearance of reverse reactions results in decomposition onset of secondary phases to their elementary phases. It occurs when the sum of chemical potentials exceeds the FE. In terms of interaction particles, the chemical potentials reflect the reservoirs for atoms that are involved in creating the substance and their values are also defined by experimental conditions. Naturally, at certain temperatures and pressures the chemical potentials of the same primary phases entering into different compounds have equal values.

At the same time these cohesive/formation energies for bixbyite $In_2O_3$, cassiterite $SnO_2$, ε-polymorph phase of $Ga_2O_3$ (orthorhombic phase) and molecular oxygen can be determined by the first-principles DFT approach as a difference between the total intrinsic electronic energies of indicated oxides and elementary phases participating in their formation i.e. In metal (space group 139, I4/mmm), Sn metal (space group 141, I4(1)/amd), Ga metal (space group 64, Cmca) and $O_2$ molecule, respectively. In the latter case we used the results obtained in [21] for atomization energy of $O_2$ carried out within DFT with PBE hybrid functional. Our approach based on Eqs. (1-5) allows one to calculate formation energies of defects without introducing in the model of so-called “metal rich” / “oxygen poor” and “metal poor” / “oxygen rich” conditions imposed on In and O chemical potentials, respectively. Such approach results in non-physically wide range of formation energies (see Ref. [22], where FE of an O-vacancy is ranged from -0.49 eV to +2.74 eV depending on the metal/oxygen limits). In fact, it introduces an empirical uncertainty associated with the "technological" factor which is beyond the first-principles calculations.

Formation energies of metal oxides and defects calculated in this work employing QE simulation package and hybrid Perdew-Burke-Ernzerhof functional for $O_2$ molecules [21] is presented in Tables 1 and 2, respectively.

**<Table 1>**

Data in parentheses present the experimental standard enthalpies of formation [23-27] for the given substances. Small deviation between theoretical and experimental results could be attributed to a temperature difference between theory ($T$ = 0 K) and experimental conditions ($T$ = 298.15 K).

**<Table 2>**

Our value for FE of $V_O$, which is one of the most extensively studied defects in $In_2O_3$, is in accordance with those reported by Tanaka et al. [28] (1.53 eV) and by Agoston et al. [29] (1.2 eV). However, their FE values were obtained in strongly different O-rich and O-poor conditions, respectively. Since these limiting conditions were necessary attributes in their models of a defect formation, reconsideration of the characteristic value for oxygen chemical potential indicated in Refs. [28, 29] may be in order.

Besides the FEs of defects with different composition, Table 2 contains the Bader partial charges for impurity and host atoms. Bader analysis demonstrates that atoms initially introduced as neutral transited to charged states after DFT calculation. Interestingly, the partial charges of host atoms are far from values of pure ionic states i.e. $In^{3+}$ and $O^{2-}$ determined by chemical formula. The Bader charges together with the analysis of band diagrams and the position of the Fermi level allow to confirm the electronic donor or acceptor nature of these impurities. The position of Fermi level, discussing in Section 5 below, provide an additional confirmation of donor/acceptor nature of the defects.

As one can see from Table 2 there is a correlation between the highest FE and lowest reducing/oxidation states in the case of single Ga/O interstitials. Probably, this correlation is related to large local distortions of the lattice due to additional point charge and under-coordinated orbitals of the defect. Bond length changes in the 1-st atomic shell of these interstitials and increase of a unit cell volume indirectly confirm this assumption (see Table 3).

Comparative analysis of defect FEs allows separating them into several groups depending on their values:

- defects with negative or close to zero FE≤0; such defects consist of Sn or Ga substitutional atoms in the lattice sites *b* or *d*, they are formed either from single atoms (single substitution) or pairs of atoms of the same type (double substitution) or different type (mixed-type substitution); such FE indicate that formation of defect is spontaneous while the defects listed below require activation by external energy and in real conditions this can be achieved by temperature activation during the material growth;

- defects with low activation energies 0<FE<1 eV; such defects consist the pair of metal atoms in lattice sites *b* and *d* and third O or Ga atom located in SV vicinity;

- defects with moderate activation energies 1 eV<FE<2 eV; among these defects are $Sn_b$-$Ga_i$-2$O_i$, $Sn_b$-$Ga_i$ and $Ga_i$-$O_i$, i.e. defects consist of 2 or 3 filled SVs; in terms of FE magnitude these defects are also ordinary O-vacancies with FE=1.55 eV;

- defects with high activation energies FE>2 eV; they are represented as single SVs filled by O or by Ga atoms as well as Frenkel-type defect $Ga_i$-$V_O$; their formation are the most energetically unfavourable.

### 3.2. Bonds configuration

The information about bonds configuration around the defects in $In_2O_3$ is presented in Table 3. Analyzing Table 3, the following features in the defect geometry can be revealed:

- Sn or Ga substitution results in slight bond contraction with O atoms located in the 1-st shell; the contraction is practically the same for both single and double substitution; in the case of Sn atoms it is about 3%; in the mixed case or double Ga substitution such relative distance changes to around 7%;
- Introduction of a single O atom in SV together with double substitution of Sn atoms results in $O_i$ shift toward Sn by ~6-10% from the geometrical center of SV; mixed or double Ga substitution results in $O_i$ shifts from *b*-site towards *d*-site with relative displacement from the geometrical center by ~ 35% and 20%, respectively, wherein the contraction between metal atoms and the 1-st O shell are within 2-13%;
- introduction of a single Ga atom in SV results in $Ga_i$ relative shift ~17% from $In_b$ and $In_d$ atoms towards O nodes; $Sn_b$-$Ga_i$ complex also demonstrates $Ga_i$ shift from $In_d$ towards O nodes (~40%) and additionally it is characterized by $In_d$ atom shift towards $Sn_d$ with corresponding bond contraction ~20%; the latter case is also characterized by the largest change in the unit cell volume (~4%); $Sn_b$-$Ga_i$ complex with additional 2 $O_i$ atoms positioned in neighbor SVs retains the specified behavior although with smaller $Sn_b$-$In_d$ bond contraction;

- the defect totally constituted from 3 Ga atoms ($Ga_b$-$Ga_d$-$Ga_i$) is also characterized by $Ga_i$ shift from *b*- and *d*-sites towards O nodes; its displacement is almost the same as for the single $Ga_i$ defect.

**<Table 3>**

It can be concluded that substitutional atoms weakly change the bond lengths and unit cell volume (within several %). The greatest distortions are observed in the case of interstitial Ga atoms. However, the real change in bond lengths near the defect does not exceed 12%. The highest enlargement of the unit cell volume can be explained quantitatively through the increase of $Sn_b$ and $Ga_i$ effective sizes due to their low oxidation states $Sn^{+1.0}$ and $Ga^{+1.34}$ according to the Bader partitioning. Stronger shifts revealed for some atomic positions which are related to interstitial atoms in SV only. The latter is a result of the rather free movement of interstitial atom within the SV cavity due to a lower coordination number and bond strength with nearest environment.

Introduction of interstitial Ga defects in $In_2O_3$ leads to a crystal lattice disorder. We can presume that the disorder induced by the rattling nature of interstitial atoms could strongly reduce the thermal conductivity of doped $In_2O_3$. Recent experimental results by Liu et al. [6] on the significant drop of a thermal conductivity in Ga-doped $In_2O_3$ confirm this prediction. Similar reduction of the thermal conductivity was found in other thermoelectric materials [30-34].

### 3.3. Electronic band structure

In our electronic calculations we have used the following valence configurations for atoms: O $2s^2 2p^4$, In $4d^{10}5s^2 5p^1$, Sn $4d^{10}5s^2 5p^2$ and Ga $3d^{10}4s^2 4p^1$. Figure 2 shows calculated electronic band structure along $\Gamma(0,0,0)$-H$(1/2\,,-1/2\,,1/2)$-N$(0,0,1/2)$-$\Gamma$-P$(1/4\,,1/4\,,1/4)$ Brillouin zone path and the orbital-specific LDOS of pure $In_2O_3$.

**<Figure 2>**

Our calculated fundamental band gap is 1.03 eV. Although this value is in a good accordance with other GGA-PBE calculations [35-37] it is much smaller than the values 2.7-3.2 eV obtained from experiments [38-44] and from more accurate DFT models with hybrid functionals [45, 46] or empirical Coulomb interaction parameters (DFT+U) [47]. At the same time our optical band gap is by 0.76 eV larger than fundamental gap. This difference (denoted with gray arrows in Figure 2) is close to values 0.62 eV [39] and 0.8 eV [48] obtained from the optical absorption spectroscopy measurements and DFT+U calculations. The band structure details can be further analyzed from the LDOS calculations resolved for different orbitals. The upper part of the valence band (VB) shows a small dispersion and it is dominated by O 2p states with a weak overlap with In 4d and In 5p states. In the lowermost part of the first conduction band (CB) the In 5s states prevail. These states are characterized by a parabolic dispersion with an isotropic effective mass $m_x^* = m_y^* = m_z^* = 0.17m_0$. For higher energies up to 5 eV with respect to CB minimum a strong hybridization between In 5s and O 2p states is found.

Band structures and orbital-dependent DOS for $In_2O_3$ with different point defects are presented in Figures 3-5. The introduction of single $Ga_{In(b)}$ and $Sn_{In(b)}$ substitutions only weakly affects the band structure dispersion in comparison with pure $In_2O_3$. The major difference is the position of Fermi level, while for $Ga_{In(b)}$ case it remains inside the gap close to its center, for $Sn_{In(b)}$ it has climbed up to 1.76 eV above the bottom of CB (see Figure 3(b)) turning the corresponding material into an n-type degenerate semiconductor. The introduction of an O vacancy (Figure 3(a)) or an interstitial Ga (Figure 3(c)) changes the band structure somewhat similarly, it splits the first CB into two subbands with a minimum energy gap in-between ~1 eV at point $\mathrm{N}(0,0,1/2)$ of the Brillouin zone. For both these defects the Fermi level lies above the first subband manifesting strong donor behaviour. The LDOS of the first conduction subband revealed an overlap between oxygen and indium electronic orbitals with an additional mixing with Ga 4p and Ga 4s states in case of $Ga_i$ defect. The second conduction subband is low-dispersive with minimum at $\mathrm{N}(0,0,1/2)$ characterized by heavy electrons and by a pronounced peak in DOS. The LDOS analysis of this subband shows a strong mixing between O 2p, O 2s, In 5p and In 5s electronic states, again with some additional mixing with Ga 4p and Ga 4s orbitals for interstitial Ga defect.

**<Figure 3>**

The band structures of $In_2O_3$ with point defect complexes consisting of substitutions (Sn, Ga) and interstitials (Ga, O) in different combinations are more complicated (see Figures 4-5). However, we conclude that some characteristic features described above for one-site defects are also found for defect complexes. Additionally, the introduction of oxygen interstitials is accompanied by appearance of numerous energy levels inside the band gap dominated by O 2p electronic states. Also, the largest deviation in electron effective masses of the first CB is found for $Ga_i$-$O_i$ ($m_x^* = m_z^* = 0.15m_0, m_y^* = 0.14m_0$) and $Ga_{In(b)}$-$Ga_{In(d)}$-$O_i$ ($m_x^* = 0.20m_0, m_y^* = m_z^* = 0.19m_0$) defect complexes. The electron effective masses for some distinctive cases of $In_2O_3$ with defect complexes along the selected symmetries are presented in Table 4.

**<Figure 4>**

**<Figure 5>**

**<Table 4>**

In general, band structure calculations of $In_2O_3$ with point defects revealed a large donor effect from O vacancies, Sn substitutions and Ga interstitials. The position of Fermi energy in these cases indicates an n-type conductivity behavior with a strong degeneration of CB. The interstitial O on the contrary behaves as a pronounced acceptor. The calculated Bader partial charges of impurity atoms and changes in partial charges of host atoms additionally confirm n- or p-type behaviour of the defects. The charge gain obtained due to substitutional or interstitial atoms does not compensate the charge changes of surrounding atoms of the defect complex. This charge gain is the origin of CB electrons and band gap states, the latter was found to be characteristic to oxygen interstitials. For n-type behaviour our analysis of the Bader charges, consisted in comparison of charge deviation of defect from the initial local host charge against the charge of all host atoms of unit cell, demonstrate their coincidence and rather good uniformity of redistributed charge throughout the crystal. This is an indicator of non-localized nature of electrons. At the same time the maximum of the electron density distribution is in the vicinity of metallic atoms of the host lattice. Thus, the Bader analysis shows that both Sn substitutions and Ga interstitials lead to n-type conduction, while O interstitials results in p-type conduction.

We have checked the consistency of our calculations in the case of single $Sn_b$ substitution, using a relation between electron concentration $n_{CB}$ in CB and defect concentration. As it follows from semiconductor statistics $n_{CB}$ is a function of Fermi energy and CB effective mass. Single $Sn_b$ substitution corresponds to doping of 6.25%, i.e. Sn concentration $n_{Sn}$ = $1.93 \cdot 10^{21}$ $cm^{-3}$. From our calculations we have obtained the following parameters of the CB: Fermi energy with respect to the CB $E_F$-$E_C$=1.76 eV, electron CB effective mass $m_C^* = 0.17m_0$, CB charge gain due to the $Sn_b$ charge loss during the In substitution $\Delta q_{CB} = 2.35e - 1.86e \cong 0.5e$ (see Table 2). For degenerated semiconductor $n_{CB}$ at $T$ = 0 K is given by [49]:

$$n_{CB} = \frac{8\pi}{3}\left(\frac{2m_C^*(E_F - E_C)}{h^2}\right)^{3/2}. \qquad (6)$$

Using Eq. (6) we estimate CB electron concentration as $n_{CB}$ = $7.4 \cdot 10^{20}$ $cm^{-3}$. Taking into account the released charge from $Sn_b$ impurity, one can obtain the "recalculated" Sn concentration as $1.48 \cdot 10^{21}$ $cm^{-3}$ that is in reasonable accordance with aforementioned value of $n_{Sn}$.

## 4. Conclusions

Within density functional theory approach the various point defect complexes in cubic $In_2O_3$ structure have been systematically studied. The considered defects were comprised from one to four atoms of Sn, Ga and O in substitutional and interstitial (structural vacancy) positions. The formation energies, Bader partial charges, bond configurations and electronic band structures were calculated and analyzed. The energies of defect formation were ranged from -0.7 eV to more than 4 eV, depending on the number of constituent atoms and their positions. It was demonstrated that defect formation energy can be calculated without introducing any ambiguous so-called technological factors using chemical potential of atomic oxygen determined from the formation energy of its molecular form. Substitutional Ga and Sn defects are the most energetically preferable and their formations are spontaneous, while Ga double-substitution complex has the lowest formation energy. Their isovalent and donating behaviour is explained within the traditional valence concept of substitutional doping.

The formation of interstitial atoms and their complexes requires an activation energy which usually decreases with the number of atoms comprising the complex. It was shown that interstitial Ga defects demonstrate donor-like behaviour and result in degenerated conduction band accompanied by its splitting into two subbands with light and heavy electrons. Contrariwise, interstitial O defects act as acceptors and lead to the formation of acceptor levels or subbands inside the band gap. These findings can shed light on unusually strong increase of electrical conductivity in $In_2O_3$ doped by Ga.

It was also revealed the decrease of CB effective mass for interstitial Ga defects and corresponding increase of CB effective mass for interstitial O defects. The calculation of electron local density of states for conduction subbands demonstrated an overlap between O and In electronic orbitals with an additional mixing with Ga 4p and Ga 4s states in the case of $Ga_i$ defects. The obtained results are important for the interpretation of transport phenomena in $In_2O_3$ with different substitutional and interstitial defects. Physical properties of such defects could not be accurately described in the framework of the standard doping concept and density functional theory is required for both qualitative and quantitative predictions.

AUTHOR INFORMATION

**Corresponding Author**

*E-mail: dlnika@yahoo.com

Tel.: +373 790 33725

ORCID Denis Nika: https://orcid.org/0000-0002-3082-3118

**Author Contributions**

‡These authors contributed equally to this work.

**Acknowledgment**

Authors acknowledge financial support from the Moldova State project #15.817.02.29F. A.I.C. acknowledges support under Moldova State project for young scientists #19.80012.02.13F.

## Figure Captions

**Figure 1**. Schematic view of $In_2O_3$ lattice sites in the vicinity of *b*-site that is one of the corners of primitive unit cell. Blue nodes indicate In(*b*-site), yellow nodes show In(*d*-site), red nodes correspond to O(*e*-site), while gray sites indicate structural vacancy (*c*-site).

**Figure 2**. Defect-free $In_2O_3$ electronic band structure and LDOS with contribution from different orbitals. Fermi level is placed at 0 eV (dashed line).

**Figure 3**. Electronic band structure and LDOS of indium oxide with one-site point defects: (a) $V_O$, (b) $Sn_{In(b)}$ and (c) $Ga_i$.

**Figure 4**. Electronic band structure and LDOS of indium oxide with two-site point defects: (a) $Ga_i$-$O_i$ and (b) $Sn_{In(b)}$-$Ga_i$.

**Figure 5**. Electronic band structure and LDOS of indium oxide with three- and four-site point defects: (a) $Sn_{In(b)}$-$Ga_{In(b)}$-$O_i$, (b) $Sn_{In(b)}$-$Ga_i$-$2O_i$ and (c) $Ga_{In(b)}$-$Ga_{In(d)}$-$Ga_i$.

**Table 1**. Formation energy.

| Material | Formation energy (standard enthalpy of formation), eV |
|---|---|
| $In_2O_3$ | -10.00 (-9.47) |
| $SnO_2$ | -5.77 (-5.98) |
| monoclinic β-$Ga_2O_3$<br>appears at T>800 $^oC$ | -10.78 (-11.28) |
| orthorhombic ε-$Ga_2O_3$<br>most stable up to 800 $^oC$ | -9.44 (N.A.) |
| Molecular oxygen | -5.29 (-5.11) [21, 23] |

**Table 2**. Defect formation energies.

| Type of defect | Formation energy, eV | Formation energy, eV per atom | Bader charge of impurity atoms, electron charge | Average Bader charge of host atoms, electron charge |
|---|---|---|---|---|
| $V_O$ | 1.55 | 1.55 | | In=+1.78 O=-1.24 |
| $Sn_b$ | -0.17 | -0.17 | +2.35 | In=+1.83 O=-1.24 |
| $Sn_d$ | -0.11 | -0.11 | +2.32 | In=+1.83 O=-1.24 |
| $Sn_b$-$Sn_d$ | 0.47 | 0.235 | +2.28; +2.25 | In=+1.81 O=-1.24 |
| $Sn_b$-$Sn_d$-$O_i$ | 1.56 | 0.52 | +2.38; +2.40; -1.21 | In=+1.86 O=-1.23 |
| $Ga_b$ | -0.39 | -0.39 | +1.90 | In=+1.86 O=-1.24 |
| $Ga_d$ | -0.28 | -0.28 | +1.88 | In=+1.86 O=-1.24 |
| $Ga_b$-$Ga_d$ | -0.71 | -0.36 | +1.90; +1.88 | In=+1.86 O=-1.24 |
| $Ga_b$-$Ga_d$-$Ga_i$ | 2.43 | 0.81 | +1.79; +1.80; +1.02 | In=+1.79 O=-1.24 |
| $Ga_b$-$Ga_d$-$O_i$ | 2.78 | 0.93 | +1.89; +1.90; -0.67 | In=+1.86 O=-1.21 |
| $Ga_i$-$O_i$ | 3.95 | 1.98 | +1.72; -1.22 | In=+1.83 O=-1.23 |
| $Sn_b$-$Ga_d$ | -0.40 | -0.20 | +2.35; +1.86 | In=+1.83 O=-1.24 |
| $Sn_b$-$Ga_d$-$O_i$ | 2.86 | 0.95 | +2.40; +1.90; -1.01 | In=+1.86 O=-1.22 |
| $Sn_b$-$Ga_i$ | 3.73 | 1.87 | +1.00; +1.34 | In=+1.81 O=-1.23 |
| $Sn_b$-$Ga_i$-2$O_i$ | 5.69 | 1.42 | +2.37; +1.79; -1.20; -1.21 | In=+1.84 O=-1.23 |
| $O_i$ | 4.58 | 4.58 | -0.94 | In=+1.87 O=-1.21 |
| $Ga_i$ | 3.65 | 3.65 | +1.01 | In=+1.79 O=-1.23 |
| $Ga_i$-$V_O$ | 4.67 | 2.34 | +0.94 | In=+1.71 O=-1.23 |

*Three of processes are highlighted (in blue) as the most energetically unfavourable

*$In_2O_3$ Bader charges: In = +1.86*e*; O = -1.24*e*

**Table 3**. Bond lengths and unit cell volume.

| Defect type | Neighbor pair | Bond length, Å | $\Delta L/L_{In2O3}$, % | $V_{cell}$, $Å^3$ | $\Delta V/V_{In2O3}$, % |
|---|---|---|---|---|---|
| pure $In_2O_3$ | In(b) - O | 2.18 (6) | | 523.89 | |
| | In(d) - O | 2.14 ÷ 2.23 (6) | | | |
| | In(b) - In(d) | 3.81 (6) | | | |
| | In(b) - V(i) | 2.35 (1) | | | |
| | In(d) - V(i) | 2.33 (3) | | | |
| | O - V(i) | 2.36 ÷ 2.42 (6) | | | |
| | V(i) - V(i) | 3.59 | | | |
| $V_O$ | | | | 524.78 | +0.17 |
| $Sn_{In(b)}$ | Sn(b) - O | 2.11 (6) | -3.21 | 527.05 | +0.60 |
| $Sn_{In(d)}$ | Sn(d) - O | 2.08 ÷ 2.16 (6) | -2.80 ÷ -2.70 | 527.01 | +0.595 |
| $Sn_{In(b)}$-$Sn_{In(d)}$ | Sn(b) - O | 2.12 ÷ 2.16 (6) | -2.75 ÷ -0.92 | 530.51 | +1.26 |
| | Sn(d) - O | 2.09 ÷ 2.17 (6) | -2.33 ÷ -2.69 | | |
| $Sn_{In(b)}$-$Sn_{In(d)}$-$O_i$ | Sn(b) - O | 2.12 ÷ 2.21 (6) | -2.75 ÷ +1.38 | 526.54 | +0.51 |
| | Sn(d) - O | 2.09 ÷ 2.26 (6) | -2.33 ÷ +1.35 | | |
| | Sn(b) - O(i) | 2.21 (1) | -5.96 | | |
| | Sn(d) - O(i) | 2.09 (1) | -10.3 | | |
| | In(d) - O(i) | 2.18 (1) | -6.44 | | |
| $Ga_{In(b)}$ | Ga(b) - O | 2.03 (6) | -6.88 | 515.23 | -1.65 |
| $Ga_{In(d)}$ | Ga(d) - O | 1.99 ÷ 2.12 (6) | -7.01 ÷ -4.93 | 515.48 | -1.61 |
| $Ga_{In(b)}$-$Ga_{In(d)}$ | Ga(b) - O | 2.01 ÷ 2.09 (6) | -7.80 ÷ -4.13 | 506.93 | -3.24 |
| | Ga(d) - O | 1.97 ÷ 2.12 (6) | -7.94 ÷ -4.93 | | |
| $Ga_{In(b)}$-$Ga_{In(d)}$-$Ga_i$ | Ga(b) - O | 1.89 ÷ 2.22 (5) | -13.3 ÷ +1.83 | 532.27 | +1.60 |
| | Ga(d) - O | 1.89 ÷ 2.08 (5) | -11.68 ÷ -6.73 | | |
| | Ga(b) - Ga(i) | 2.78 (1) | +18.3 | | |
| | Ga(d) - Ga(i) | 2.63 (1) | +12.88 | | |
| | In(d) - Ga(i) | 2.61 (1); 2.71 (1) | +12.02; +16.31 | | |
| $Ga_{In(b)}$-$Ga_{In(d)}$-$O_i$ | Ga(b) - O | 1.99 ÷ 2.17 (6) | -8.72 ÷ -0.46 | 515.31 | -1.64 |
| | Ga(d) - O | 1.91 ÷ 2.26 (6) | -10.75 ÷ +1.35 | | |
| | Ga(b) - O(i) | 3.28 (1) | +39.6 | | |
| | Ga(d) - O(i) | 1.91 (1) | -18.0 | | |
| | In(d) - O(i) | 2.24 (1); 2.81 (1) | -3.86; +20.6 | | |
| $Ga_i$-$O_i$ | O - Ga(i) | 1.85 ÷ 2.17 (5) | -21.6 ÷ -10.3 | 543.72 | +3.79 |
| | Ga(i) - O(i) | 3.40 (1) | -5.29 | | |
| | In(b) - Ga(i) | 2.72 (1) | +15.74 | | |
| | In(d) - Ga(i) | 2.73 ÷ 2.82 (3) | +17.17 ÷ +16.53 | | |
| | In(b) - O(i) | 2.29 (1) | -2.55 | | |
| | In(d) - O(i) | 1.69 (1); 2.07 (1) | -27.47; -11.16 | | |
| $Sn_{In(b)}$-$Ga_{In(d)}$ | Sn(b) - O | 2.09 ÷ 2.13 (6) | -4.13 ÷ -2.29 | 518.73 | -0.98 |
| | Ga(d) - O | 1.98 ÷ 2.11 (6) | -7.48 ÷ -4.48 | | |
| $Sn_{In(b)}$-$Ga_{In(d)}$-$O_i$ | Sn(b) - O | 2.07 ÷ 2.14 (6) | -5.05 ÷ -1.83 | 521.49 | -0.46 |
| | Ga(d) - O | 1.90 ÷ 2.11 (6) | -11.21 ÷ -5.38 | | |
| | Sn(b) - O(i) | 3.04 (1) | +29.4 | | |
| | Ga(d) - O(i) | 1.90 (1) | -18.45 | | |
| | In(d) - O(i) | 2.19 (1); 2.33 (1) | -6.0; 0 | | |
| $Sn_{In(b)}$-$Ga_i$ | Sn(b) - O | 2.28 (3) | +4.59 | 546.48 | +4.31 |
| | Sn(b) - Ga(i) | 2.48 (1) | +5.53 | | |
| | O - Ga(i) | 1.90 (3) | -19.49 ÷ -21.49 | | |
| | Sn(b) - In(d) | 2.98 (3) | -21.78 | | |
| | In(d) - Ga(i) | 3.31 (3) | +42.06 | | |
| $Sn_{In(b)}$-$Ga_i$-$2O_i$ | Sn(b) - O | 2.03 ÷ 2.20 (6) | -6.88 ÷ +0.92 | 546.27 | +4.27 |
| | Sn(b) - Ga(i) | 2.81 (1) | +19.57 | | |
| | O - Ga(i) | 1.87 ÷ 2.06 (5) | -20.76 ÷ -14.88 | | |
| | Sn(b) - In(d) | 3.60 ÷ 3.67 (3) | -5.51 ÷ -3.67 | | |
| | In(d) - Ga(i) | 2.70 ÷ 2.81 (3) | +15.88 ÷ +20.60 | | |
| $O_i$ | In(b) - O(i) | 2.35 (1) | 0 | 529.66 | +1.10 |
| | In(d) - O(i) | 2.22 (3) | -4.72 | | |
| $Ga_i$ | O - Ga(i) | 1.98 (3) | -16.1 ÷ -18.2 | 543.00 | +3.65 |
| | In(b) - Ga(i) | 2.69 (1) | +14.47 | | |
| | In(d) - Ga(i) | 2.70 (3) | +15.88 | | |
| $Ga_i$-$V_O$ | O - Ga(i) | 1.94 ÷ 2.12 (3);<br>2.89 (1) | -17.8 ÷ -12.40;<br>+19.42 | 540.93 | +3.25 |
| | In(b) - Ga(i) | 2.82 (1) | +20.0 | | |
| | In(d) - Ga(i) | 2.65 ÷ 2.70 (3) | +13.73 ÷ +15.88 | | |

**Table 4**. Electron effective mass.

| Type of defect | Electron effective mass, $m_0$ | | | | | |
|---|---|---|---|---|---|---|
| | Γ → P | | Γ → N | | Γ → H | |
| | 1st CB | 2nd CB | 1st CB | 2nd CB | 1st CB | 2nd CB |
| pure $In_2O_3$ | 0.17 | | 0.17 | | 0.17 | |
| $Sn_b$ | 0.16 | | 0.16 | | 0.17 | |
| $Sn_b$-$Sn_d$-$O_i$ | 0.16 | | 0.16 | | 0.16 | |
| $Ga_b$-$Ga_d$ | 0.17 | | 0.17 | | 0.17 | |
| $Ga_b$-$Ga_d$-$O_i$ | 0.19 | | 0.19 | | 0.20 | |
| $Ga_i$ | 0.16 | 8.78 | 0.15 | 4.68 | 0.15 | 4.68 |
| $Ga_b$-$Ga_d$-$Ga_i$ | 0.16 | 3.51 | 0.16 | 4.68 | 0.16 | 4.26 |

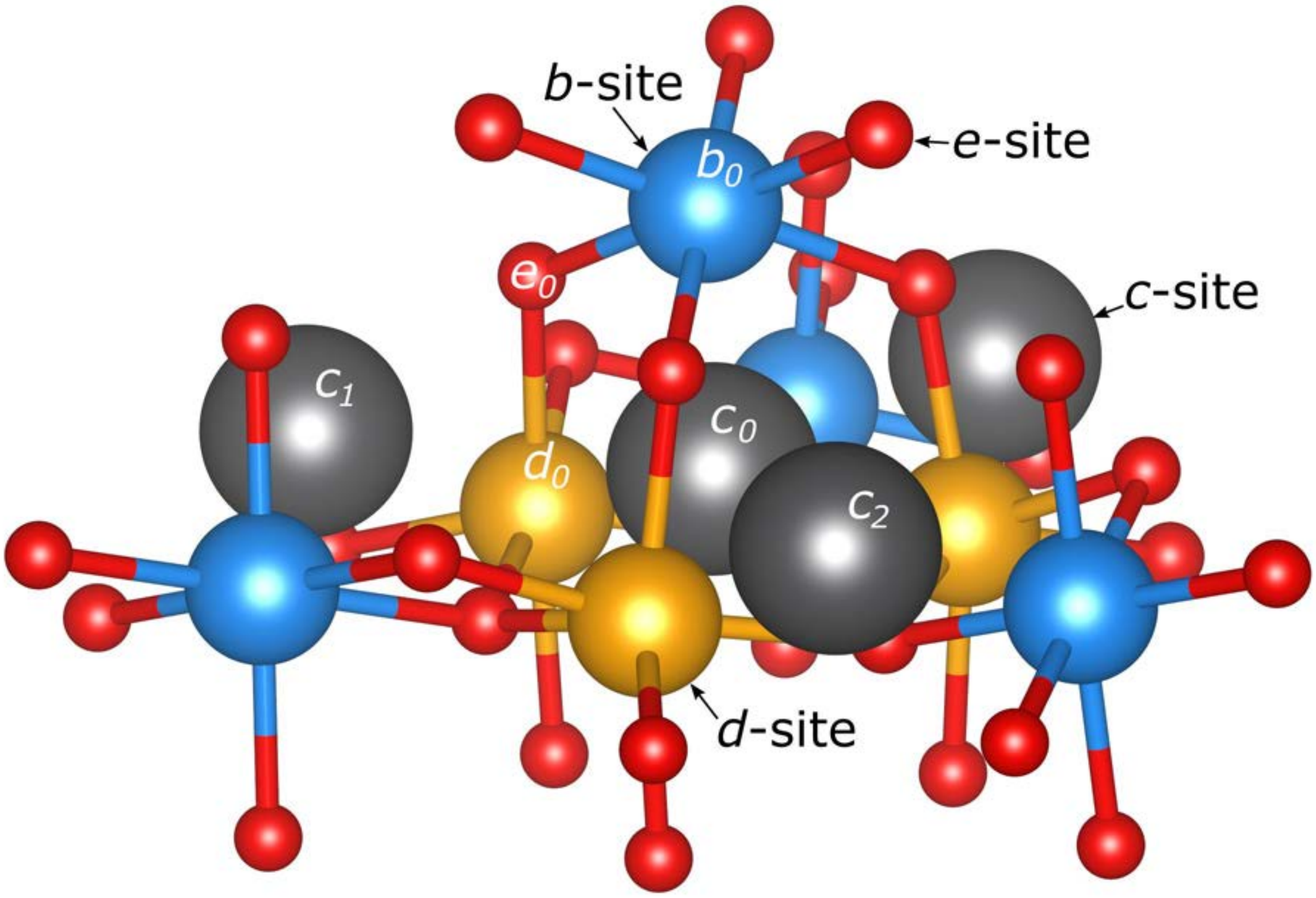

Figure 1 of 5. A. I. Cocemasov et al.

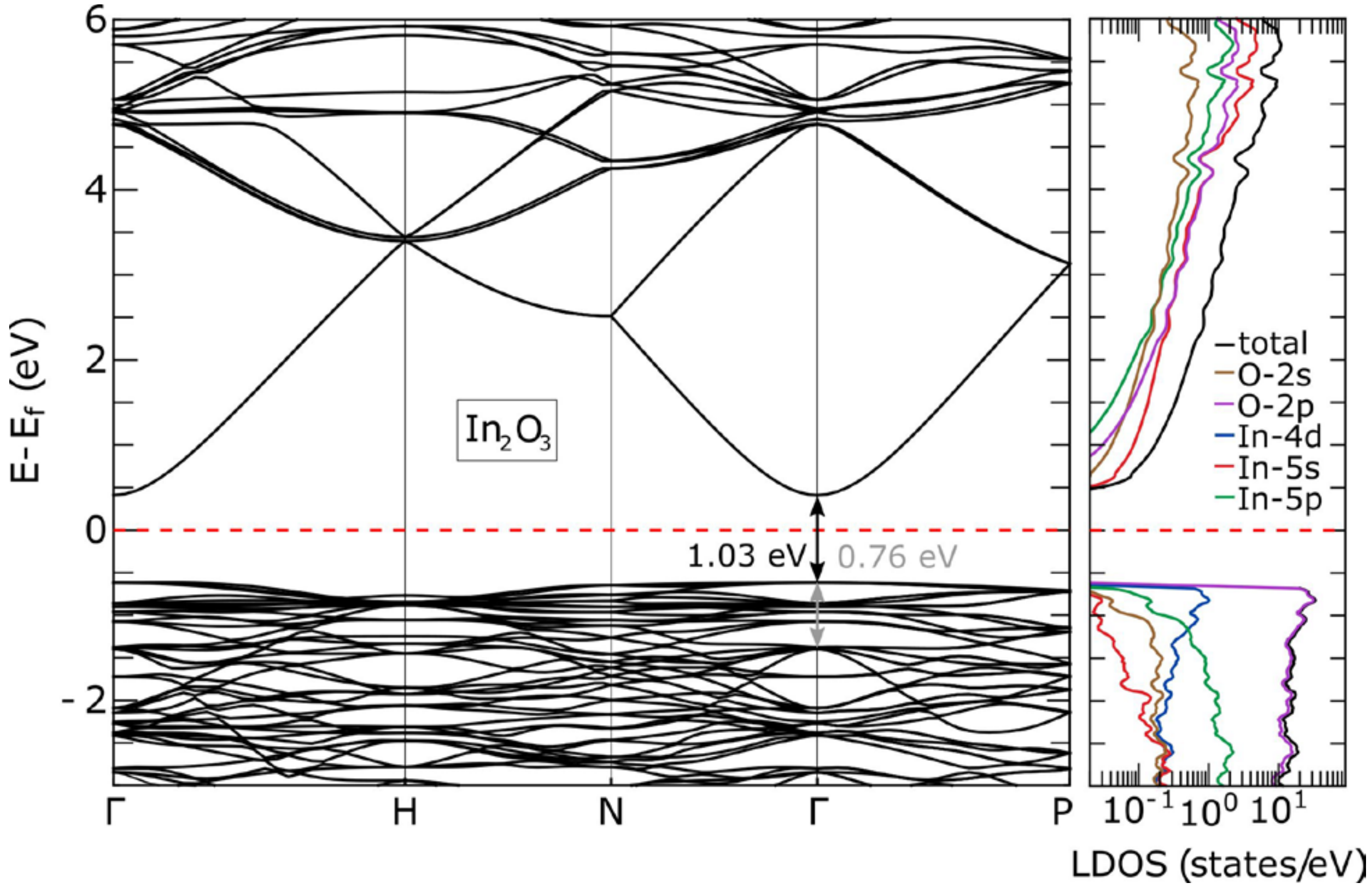

Figure 2 of 5. A. I. Cocemasov et al.

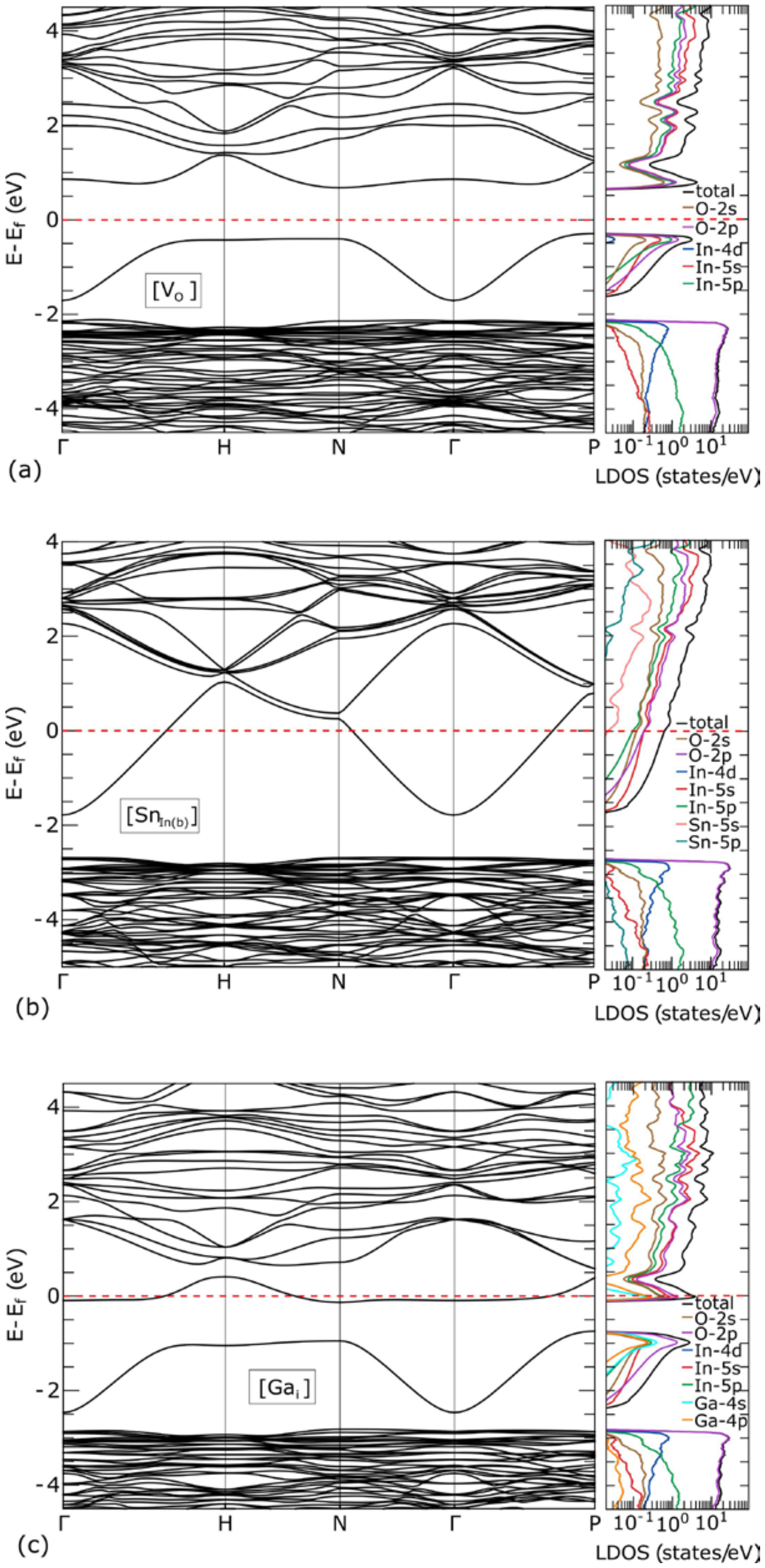

Figure 3 of 5. A. I. Cocemasov et al.

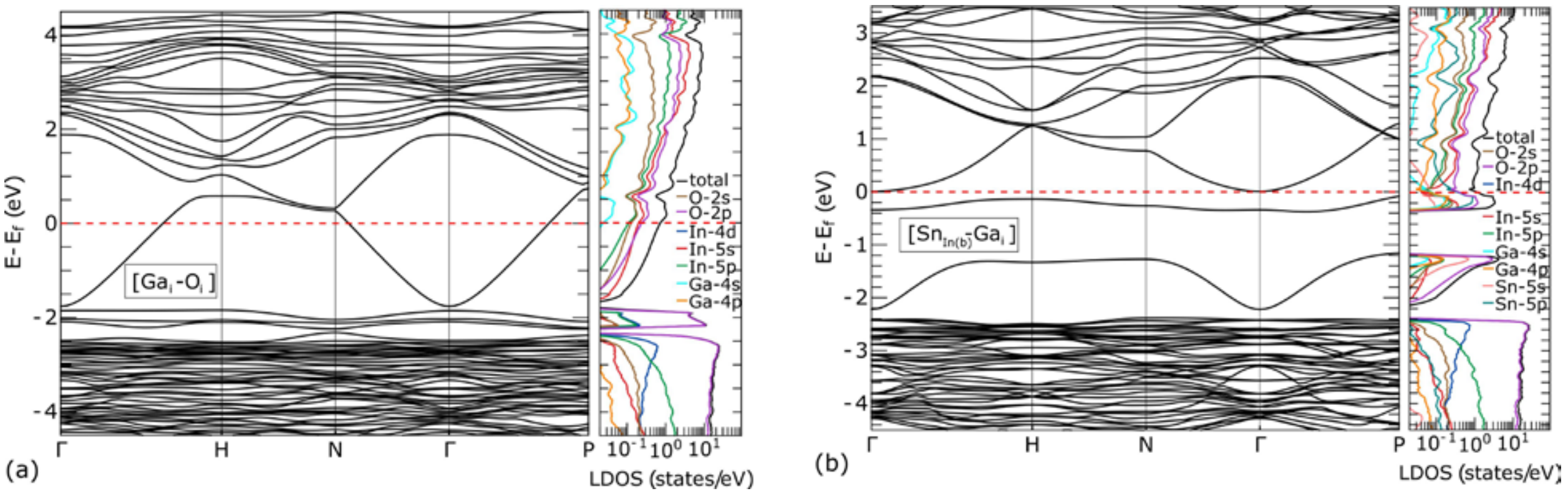

Figure 4 of 5. A. I. Cocemasov et al.

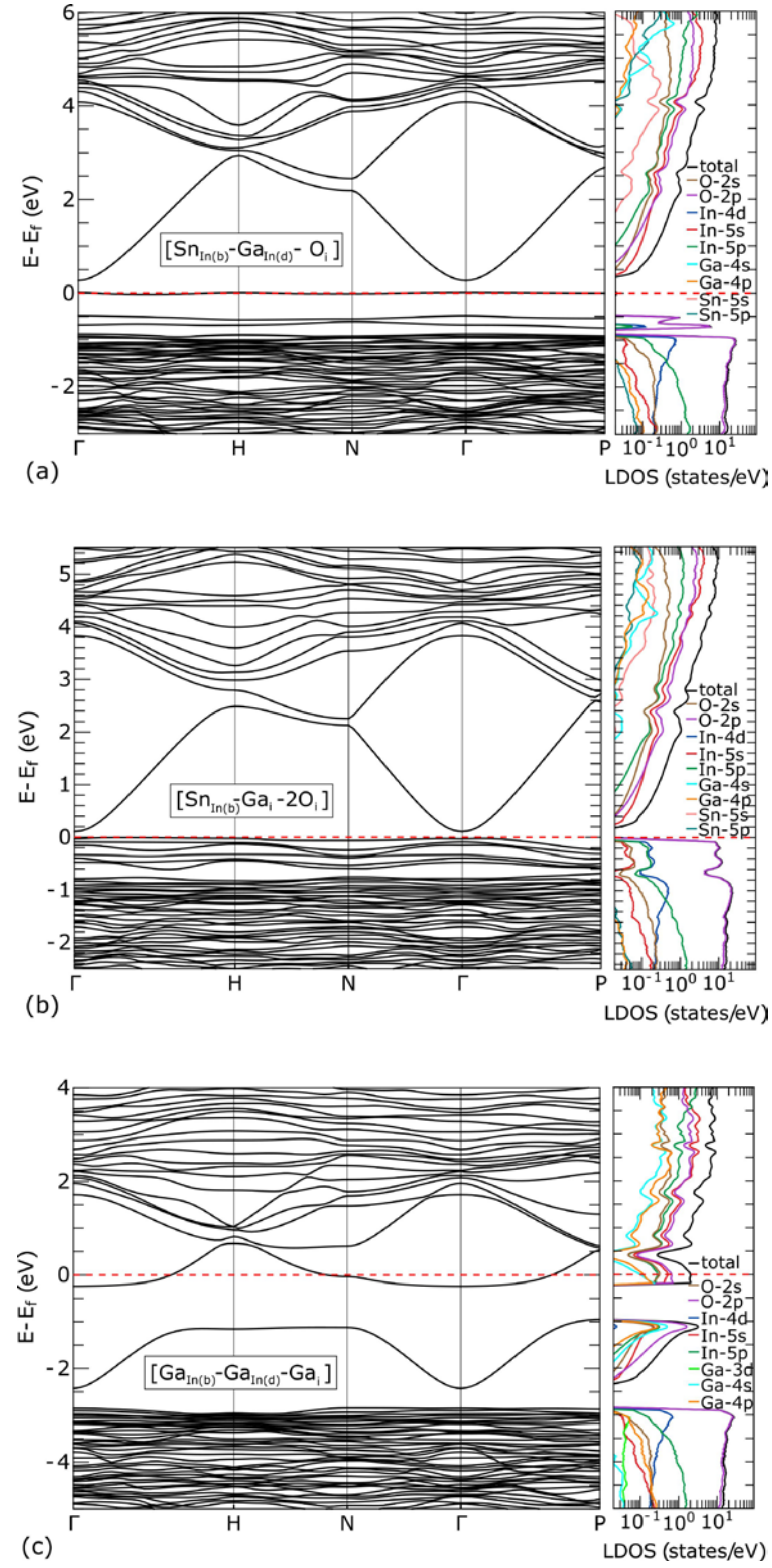

Figure 5 of 5. A. I. Cocemasov et al.